\documentclass[conference]{IEEEtran}
\IEEEoverridecommandlockouts

\usepackage{cite}
\usepackage{amsmath,amssymb,amsfonts}
\usepackage{algorithmic}
\usepackage{array}
\usepackage{graphicx}
\usepackage{tabularx}
\usepackage{textcomp}
\usepackage{xcolor}
\usepackage{comment}
\usepackage{url}
\usepackage{subfigure}
\usepackage{float}     
\usepackage{placeins} 
\usepackage{graphicx}
\usepackage{array}
\usepackage{booktabs}
\usepackage{multirow}
 
\def\BibTeX{{\rm B\kern-.05em{\sc i\kern-.025em b}\kern-.08em
    T\kern-.1667em\lower.7ex\hbox{E}\kern-.125emX}}
\begin{document}

\title{Embedding computational neurorehabilitation in clinical practice using a modular intelligent health system

\thanks{This research was partially supported by the P \& K foundation, Innosuisse Flagship SwissNeuroRehab PFFS-21-64, and SwissUniversities ORD Explore}
}

\author{
\IEEEauthorblockN{
Thomas Weikert\textsuperscript{1,2,3}, 
Eljas Roellin\textsuperscript{4}, 
Monica Pérez-Serrano\textsuperscript{1,2}, 
Elisa Du\textsuperscript{1,2},  
Lukas Heumos\textsuperscript{4},\\
Fabian J. Theis\textsuperscript{4,5}, 
Diego Paez-Granados\textsuperscript{2}, 
Chris Easthope Awai\textsuperscript{1}\\\\
}
\IEEEauthorblockA{
\textsuperscript{1}\textit{Data Analytics \& Rehabilitation Technology (DART), Lake Lucerne Institute, Switzerland} \\
\textsuperscript{2}\textit{Spinal Cord and Artificial Intelligence (SCAI) Lab, ETH Zurich and Swiss Paraplegic Research, Switzerland} \\
\textsuperscript{3}\textit{Inria, Paris-Saclay, France} \\
\textsuperscript{4}\textit{Institute of Computational Biology, Helmholtz Center Munich, Germany} \\
\textsuperscript{5}\textit{Department of Mathematics, Technical University of Munich, Germany} \\\\
}
}

\maketitle

\begin{abstract}
A significant and rising proportion of the global population suffer from non-communicable diseases, such as neurological disorders. Neurorehabilitation aims to restore function and independence of neurological patients through providing interdisciplinary therapeutic interventions. Computational neurorehabilitation, an automated simulation approach to dynamically optimize treatment effectivity, is a promising tool to ensure that each patient has the best therapy for their current status. However, computational neurorehabilitation relies on integrated data flows between clinical assessments, predictive models, and healthcare professionals. Current neurorehabilitation practice is limited by low levels of digitalization and low data interoperability. We here propose and demonstrate an embedded intelligent health system that enables detailed digital data collection in a modular fashion, real-time data flows between patients, models, and clinicians, clinical integration, and multi-context capacities, as required for computational neurorehabilitation approaches. We give an outlook on how modern exploratory data analysis tools can be integrated to facilitate model development and knowledge inference from secondary use of observational data this system collects. With this blueprint, we contribute towards the development of integrated computational neurorehabilitation workflows for clinical practice.
\end{abstract}


\section{Introduction}

The World Health Organization indicates that one in three people live with a health condition that could benefit from rehabilitation \cite{gimigliano2017rehabilitation}. This number is prognosticated to increase significantly over the coming decades, as global populations age and lifestyle changes result in increasing risk factors for non-communicable diseases (NCDs) \cite{devaux2020will, chong2024global}. One major class of NCDs is stroke, where incidence and prevalence are outpacing the NCD mean \cite{cheng2024projections}. Stroke is a leading cause of long-term disability, resulting in a growing number of stroke survivors with sensorimotor impairment. Up to 70\% of stroke survivors suffer from acute limb impairment  \cite{lawrence2001estimates}, significantly impacting their independence and quality of life \cite{markus2020tracking}. To meet the growing need for efficient and effective neurorehabilitation (NR)  in stroke survivors, current approaches must be optimized for delivery efficacy, cost-effectiveness, continuous care, and better outcomes. Integrating computational neurorehabilitation (compNR) in clinical practice promises to achieve this optimization through precision approaches tailored to the individual. These carry the promise of more effective treatments through personalization, more efficient intervention selection through data-driven clinical decision support, and more rapid and reliable intervention delivery via automation \cite{reinkensmeyer2016computational,johnson2022computational, Panch2018, french2022precision, lin2024transforming}.\\

compNR is an interdisciplinary approach that models neural plasticity and motor learning to enhance movement recovery for individuals with neurological impairments. It leverages computational models to simulate the adaptive mechanisms of the central nervous system, aiming to continuously predict the best therapeutic intervention for a given patient at each point in time \cite{reinkensmeyer2016computational}.  This concept has been implemented with various Artificial Intelligence (AI) approaches. For instance, Ye et al. \cite{ye2025ai} performed adaptive therapy planning by using reinforcement learning (RL), an approach where an agent learns to make optimal decisions through trial and error by receiving rewards or penalties based on its actions \cite{10162185}. Cotton et al. \cite{cotton2024causal} propose the use of causal models to recurrently identify optimal dynamic treatment regimens based on models of motor control and neural plasticity. Adans et al. \cite{adans2021enabling} integrate wearable sensors with traditional machine learning models to improve the prediction of recovery trajectories, indicating that continuous, high-resolution data is relevant for accurate prediction. In synthesis, there is a rapidly growing evidence base that compNR is able to predict optimal dynamic treatment regimes on an individual basis, and that this prediction can be improved by adding more granular knowledge of an individual's health status. Increasingly detailed solutions exist to digitally assess and monitor health status through multi-modal wearables, small sensors that are attached to the body and record biosignals. These sensors are increasingly used to monitor and understand motion \cite{pohl2025construct, lohse2024validation, wang2025wearable} and even influence behavior in real-time \cite{mayrhuber2024encouraging} . Wearables have been used to track stroke-specific posture \cite{pohl2022accuracy}, functional activity \cite{pohl2022classification}, and speech \cite{song2024wearable}. Emerging sensors hold promise to provide insights on other biosignals that may be highly relevant for trajectory prediction, including continuous measurement of blood pressure \cite{olawade2024integrating}, metabolites from sweat \cite{brasier2024applied}, and blood glucose \cite{wada2018outcome}. Traditional clinical assessments are being refined by wearable sensors and consumer-grade video cameras that can support more reliable clinical scoring or enable the calculation of new metrics. For instance, Werner et al. and Maceira et al. \cite{werner2022wearable, maceira2019wearable} both propose systems to support clinical scoring for rating upper limb movement quality in stroke rehabilitation. Novel metrics may be extracted from purpose-built computer vision approaches such as OpenCap \cite{uhlrich2023opencap} for walking or differentiable biomechanics of upper limb movements \cite{unger2024differentiable}. 
Taken together, computational approaches from the machine-learning domain and sensing technology are nearing a confluence where in combination they can provide fertile ground for individualized rehabilitation strategies built upon a detailed health status of the individual \cite{kosorok2019precision}. The emerging concept of digital twins extends this vision, creating virtual patient models that continuously integrate real-world data to simulate therapy effects and predict recovery trajectories \cite{katsoulakis2024digital}. 

\begin{figure*}[htbp]
    \centering
    \includegraphics[width=0.94\textwidth]{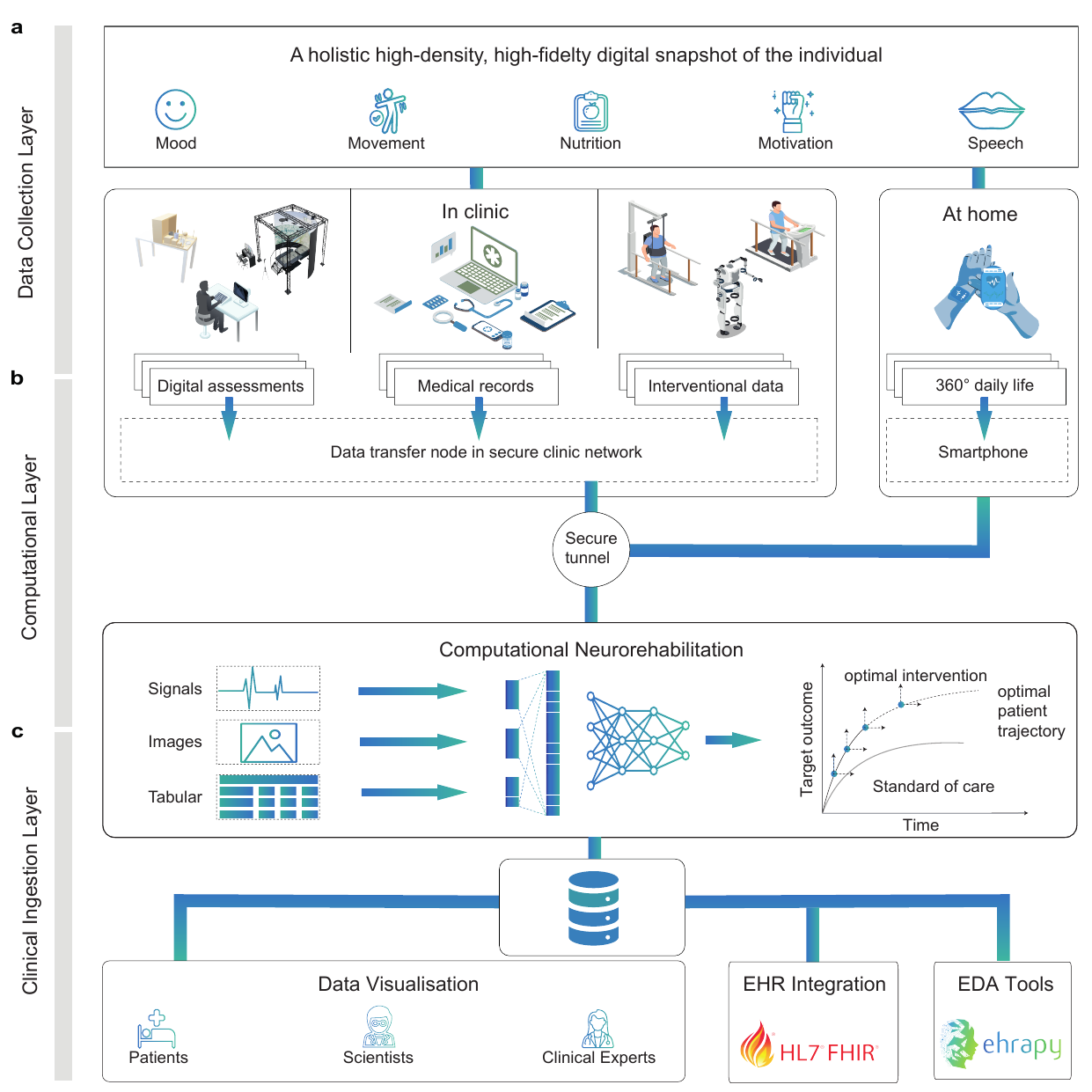}
    \caption{Schematic overview of the proposed i-health system depicting the three-layer model. (a) \textit{Data Collection} shows individual examples for clinical and home data sources to capture continuous health information. (b) \textit{Computation} shows a schematic representation of computational neurorehabilitation from data to prediction of the optimal intervention. (c) \textit{Clinical Ingestion} depicts data visualization modes and interoperability interfaces to clinical information systems and data analysis tools. }
    \label{fig:HC-setup}
\end{figure*}

While these advances bode well for the development of compNR, both Digital Twins and general AI approaches rely heavily on large high-quality, representative datasets to make accurate predictions. In general healthcare, data is often incomplete, biased, or not standardized. In NR specifically, relevant data needs to be collected across the continuum of care with diverse data collection instruments in both clinical and home contexts. These challenges are compounded through siloed data, low interoperability and comparability, and low levels of digitalization. However, to develop meaningful and generalizable models, observational data from clinical NR in volume is required. This requires changes in clinical practice and clinical data collection systems to enable the continuous integration of wearable sensors, robotic intervention platforms, and novel digital metrics of health status.

Clinical data collection systems form the backbone of data collection in clinical routine. They are used to collect all information about a patient, including demographics, diagnosis, intervention data, intervention planning, and billing. Historically, these systems are best suited to collect single-value variables in a database. The integration of continuous data-streams, such as are produced by wearables, remains an emerging topic \cite{brasier2020device}. Additionally, due to the prolonged nature of NR, data may arrive from clinical and non-clinical contexts, such as home, exacerbating the integration challenge. Furthermore, interoperability between these systems is by design not always given. Multiple initiatives are underway to improve interoperability and integration of time-series data into these platforms. They can roughly be grouped into three levels:
\textit{System-level approaches} provide a mandated space for the cross-border exchange of health data \cite{grundstrom2018health, karampela2018personal} and centralized trusted research environments (TREs). For instance, the European Health Data Space (EHDS) seeks to establish a standardized framework for cross-border health data sharing \cite{ehds2023}. It promotes AI-driven healthcare but currently focuses on conventional EHRs, neglecting high-frequency sensor data critical for NR. The Swiss Personalized Health Network (SPHN) \cite{lawrence2020sphn} and BioMedIT \cite{coman2020sphn} are among the most prominent efforts in Switzerland aimed at standardizing and improving electronic health records (EHRs), biomedical data infrastructure, and secure health data storage.
\textit{Interoperability approaches} aim to define common data models and formats to facilitate the sharing and exploitation of health data. Popular interoperability frameworks, such as HL7's Fast Health Interoperability Resources (FHIR) \cite{bender2013hl7}, the Observational Medical Outcomes Partnership (OMOP) Common Data Model (CDM) of the Observational Health Data Sciences and Informatics (OHDSI) collaborative \cite{hripcsak2015observational}, and Resource Description Framework (RDF) \cite{pan2009resource} include representations of semantic data, including medical reports, medical images, and clinical data. However, to date, there are only emerging provisions for time-series data arriving in different formats from different devices and sources \cite{9762855}. 
\textit{Mobile Data collection frameworks} aim to enable distributed data collection across clinical and home contexts using a wide range of different sensors and devices. Prominent web or mobile data collection frameworks include RedCAP \cite{harris2009research}, RADAR-Base \cite{ranjan2019radar}, and CLAID \cite{langer2024claid}. RedCAP mirrors a traditional clinical data capture system, allowing for the scheduled use of questionnaires and data entry in a web-based interface. A recently released mobile version, MyCAP, integrates with personal mobile phones and allows for user-friendly prompted data capture. However, the integration of time-series data is not currently provisioned. RADAR-Base and CLAID are both designed to be mobile-first and to capture diverse data streams from mobile devices. They enable the collection and transmission of data from sensors linked to a mobile device, prompted phone interactions, and passive mobile phone data, such as GPS. These tools, however, require technical competence to integrate and maintain, and have no link to traditional clinical data collection systems. 

To consolidate detailed health status data at scale for compNR, an intelligent health (i-health) system is needed. I-health systems enable the easy collection and integration of wearable sensor data streams, clinical data, intervention data, and digital assessment outcomes across home and clinical contexts \cite{10162185}. We here propose an i-health system that enables the continuous capture and analysis of detailed health status in a "real time" loop with patient and clinician inputs for use within standard NR settings. Our system is designed around clinical routine with extensive usability and fit-for-purpose testing in both clinical and home contexts. Through the seamless integration of many data types and sources, coupled with integration of significant bi-directional compute and extensive visualization and interoperability, we aim to both empower model development and provide a framework for the clinical implementation of compNR.

\section{Method}

Our embedded i-health system encompasses three main layers (Figure \ref{fig:HC-setup}): (1) a modular data collection layer, (2) a computational layer to calculate clinically established metrics, run compNR modules and processes data for visualization, and (3) a clinical ingestion layer, which allows no-code data exploration and ensures seamless interoperability with existing hospital infrastructure and emerging data analytic frameworks. These three layers interact continuously to create value for healthcare professionals and medical data scientists, by addressing key challenges faced by both. Next, we discuss each layer in detail, outlining an optimized framework for compNR.

\subsection{Data Collection Layer} 
The proposed i-health system is designed for the continuum of care and is hence able to collect health data across both in-person clinical, ambulatory, and home settings. Multi-modal rehabilitation data includes digital assessments, medical records, and continuous monitoring using wearable and ambient sensors. Conceptually, this data encompasses all critical aspects of the patient’s current condition. Any intervention performed to change this status is recorded as interventional data.

\subsubsection{Clinical context} 
In the clinical context, organizational motivation and user needs are two main predictors of successful implementation \cite{wen2010developing}. To respect these, a modular ecosystem of mobile apps was designed around the requirements of clinical teams. Typical prioritization included aspects related to capture time, user-friendliness, and clinical meaningfulness. Each application mirrored a typical standard clinical assessment that was already being performed as part of clinical routine. Through iterative user-testing, we created applications that received high usability ratings, reduced data acquisition and documentation time, and enhanced the data capture through wearable or ambient sensing. These apps, in combination with a wearable continuous monitoring system, ensure detailed capture of the patient objective and perceived health status. 
Intervention data is extracted from therapy plans, therapy logging, and technical data from various robotic rehabilitation devices retrieved via a universal connection framework (ROS4HC \cite{ros2hc2024}).  Medical and demographic data can be ingested in risk-mitigated form from the clinical information system if an open interface is provided by the clinic. Alternatively, an individual registration of each patient into the system is also possible.
All data collected within a clinical context is aggregated in a data transfer node (DTN), which serves as the central hub for all local tools. The DTN comes equipped with required software, managing key functions such as data encryption and file transmission. It also features a concealed wireless network, which can be integrated into the clinical network infrastructure if permitted by policy. The DTN pseudonomizes, encrypts, and transfers the data through a secure tunnel to a trusted computational environment.\

\subsubsection{Home context} 
In the home context, a central home hub and a companion application on the personal mobile device performs core services analogous to the DTN. Patient health status is captured using guided home assessments that are prompted at user-defined points in time. These include a wide span of multi-modal data based on validated home assessments. Wearable sensor data is captured using the companion app on the personal mobile phone, which enables easy interfacing with device data, connected sensors, and home devices. When in the home network, data is automatically transferred to the home hub for encryption and secure transmission. By including this approach in the design from the ground up, continuous tracking of daily behaviors and rehabilitation progress becomes possible beyond clinical settings. Similar to in-clinic applications, the development of smartphone applications follows a user-centered design approach, employing an iterative process to optimize user experience.

\subsection{Computational Layer}
Collected data is securely transferred through a tunnel to an on-premise computational environment within our research institute. Incoming raw data is stored, pre-processed, and analyzed. Each workflow follows a microservice-based architecture, utilizing multiple computing environments based on computational requirements. An orchestrator enables automation and triggers separate workflows in constant intervals. Finally, the processed data is stored in a structured data repository, ensuring interoperability for efficient distribution. Individual analytics services and multi-modal machine learning models can be applied with minimal data engineering overhead, simplifying the development of advanced health tracking and functional assessment tools for rehabilitation. Analyses such as associating patients with rehabilitation state, unsupervised patient clustering, differential comparison between patient groups, patient trajectory analysis, and more can be facilitated with the ehrapy framework \cite{heumos2024open}.

\subsection{Clinical Ingestion Layer}

The Clinical Ingestion Layer consists of two modules, the \textit{EHR Integration Module} and the \textit{Visualization Module}.
The \textit{EHR Integration Module} ensures interoperability between the proposed i-health system and existing EHR systems. The proposed i-health system provides health status data, intervention data, and prediction model output for ingestion via a secure API using industry standard formats. Clinical information systems with the ability to access our secure access tunnel, retrieve this data, and re-identify it based on the clinical key can integrate this data into their existing data models. The same API may be opened for external connections to exploratory data analysis tools, such as Ehrapy. These tools can be used to inform and refine model architectural choices and hyperparameters, iteratively improving the computational modules at the heart of compNR.

The \textit{Visualization Module} provides an HTML5 webpage that displays patient data in interactive dashboards. The dashboards are pre-configured in a user-centric manner, with a preset for "clinical perspective", "patient perspective", and "exploration perspective". Each perspective is tuned to the user needs, collected through extensive usability evaluations, and allows detailed data exploration through interaction. The underlying engine is Grafana \cite{grafana}, a web-based open observability platform. Grafana allows for rapid configuration of new data dashboards, facilitating the integration of new data representations.

\section{Results}
We have successfully embedded the described  i-health system into a member clinic of SwissNeuroRehab, a national research consortium for NR across the continuum of care. In this context, our i-health system was tailored for stroke rehabilitation, and configured to capture key components of the International Classification of Functioning, Disability, and Health (ICF) \cite{who2001icf}. 

\begin{table*}[h!]
    \centering
    \caption{Automated assessment battery embedded into clinical practice}
    \label{tab:clinical_outcomes} 
    \renewcommand{\arraystretch}{1.5} 
    \normalsize
    \resizebox{\textwidth}{!}{%
    \begin{tabular}{p{3cm} | l | p{6.cm} | p{2.5cm} | p{2cm} | p{2cm}}
        \hline
        \textbf{Clinical Outcome} & \textbf{Test} & \textbf{Measure} & \textbf{Traditional Scoring} & \textbf{Assessments per week} & \textbf{Wearable Sensors} \\
        \hline
        \multirow{1}{*}{Arm Function}  & Action Research Arm Test & Assesses grasp, grip, pinch, and gross motor movements & Trained therapist / pen \& paper & 2x & IMU, Video \\
        \hline
        \multirow{1}{*}{Activity Levels} & Physical Activity Monitoring & Quantifies physical activities in low, moderate, and vigorous times & Trained therapist & 3x & IMU \\
        \hline
        \multirow{1}{*}{Arm Use} & Physical Activity Monitoring & Quantifies the use of left and right arm within and outside of therapy & Trained therapist & 3x & IMU \\
        \hline
        \multirow{1}{*}{Linguistic Skills} & Frenchay Dysarthria Assessment (FDA) & Evaluates disorders resulting from neuromuscular impairments  & Trained therapist / pen \& paper & 3x & Tablet, Audio \\
        \hline
        \multirow{1}{*}{Linguistic Skills} & Bogenhausener Dysarthrieskalen (BoDyS) & Evaluates dysarthria in adults with neurological condition  & Trained therapist / pen \& paper & 3x & Tablet, Audio \\
        \hline
        \multirow{1}{*}{Walking Speed} & 10m Walking Test & Measures walking speed over a fixed distance (m/s) & Stopwatch & 3x & Video \\
        \hline
        \multirow{1}{*}{Perceived Fatigue} & Fatigue Severity Scale & Reports impact of fatigue on daily function & Pen \& paper & 5x & Tablet \\
        \hline
        \multirow{1}{*}{Perceived Anxiety} & Hospital Anxiety \& Depression Scale & Reports anxiety and depression in hospitalized patients and outpatient settings & Pen \& paper & 5x & Tablet \\
        \hline
        \multirow{1}{*}{Perceived Depression} & Beck Depression Inventory II Scale & Reports the severity of depressive symptoms & Pen \& paper & 5x & Tablet \\
        \hline
        \multirow{1}{*}{Perceived Sleep} & Epworth Sleepiness Scale & Reports daytime sleepiness and likelihood of falling asleep & Pen \& paper & 5x & Tablet \\
        \hline
        \multirow{1}{*}{Perceived Fatigue} & Fatigue Scale for Motor and Cognitive Functions & Reports motor and cognitive aspects of fatigue for MS patients & Pen \& paper & 5x & Tablet \\
        \hline
    \end{tabular}
    \label{tab:assessment_battery} 

    }
\end{table*}

We developed digitally enhanced versions of traditional assessments on a rolling basis to support clinical practice in their manual inspection of patients’ health status. Through interviews, operational constraints were identified, and a list of requirements for digital assessments was defined. After an initial development period, the proposed solution was implemented in a limited test period in which user feedback was continuously solicited. Until convergence (between 1-3 development cycles), this pattern was repeated. Over the course of the implementation we assembled a battery of assessments covering movement and speech therapy, coupled with patient reported outcomes (Table \ref{tab:assessment_battery}). We provided tablet applications to the physiotherapists and wearable sensors. Finally, we installed a DTN within the clinical network in close collaboration with the clinic IT to ensure that all applications were able to send data to the DTN. 
 
The computational layer consists of multiple servers and networks, each serving a specific function. A landing zone acts as the entry point for packaged and encrypted node data from a designated DTN. Upon arrival, the data is decrypted and stored in object storage as individual blocks. An orchestrator then automatically notifies relevant analytics services when compatible data is available for processing. These services run in Dockerized containers, ensuring easy replication and secure operation. Once processing is complete, the orchestrator triggers two actions: an internal visualization service updates web-based dashboards accessible within secure networks, and an export service packages the processed data using industry-standard health protocols, returning it to the donating clinic for ingestion.

\begin{figure*}[t]
    \centering
    \includegraphics[width=0.95\textwidth,keepaspectratio]{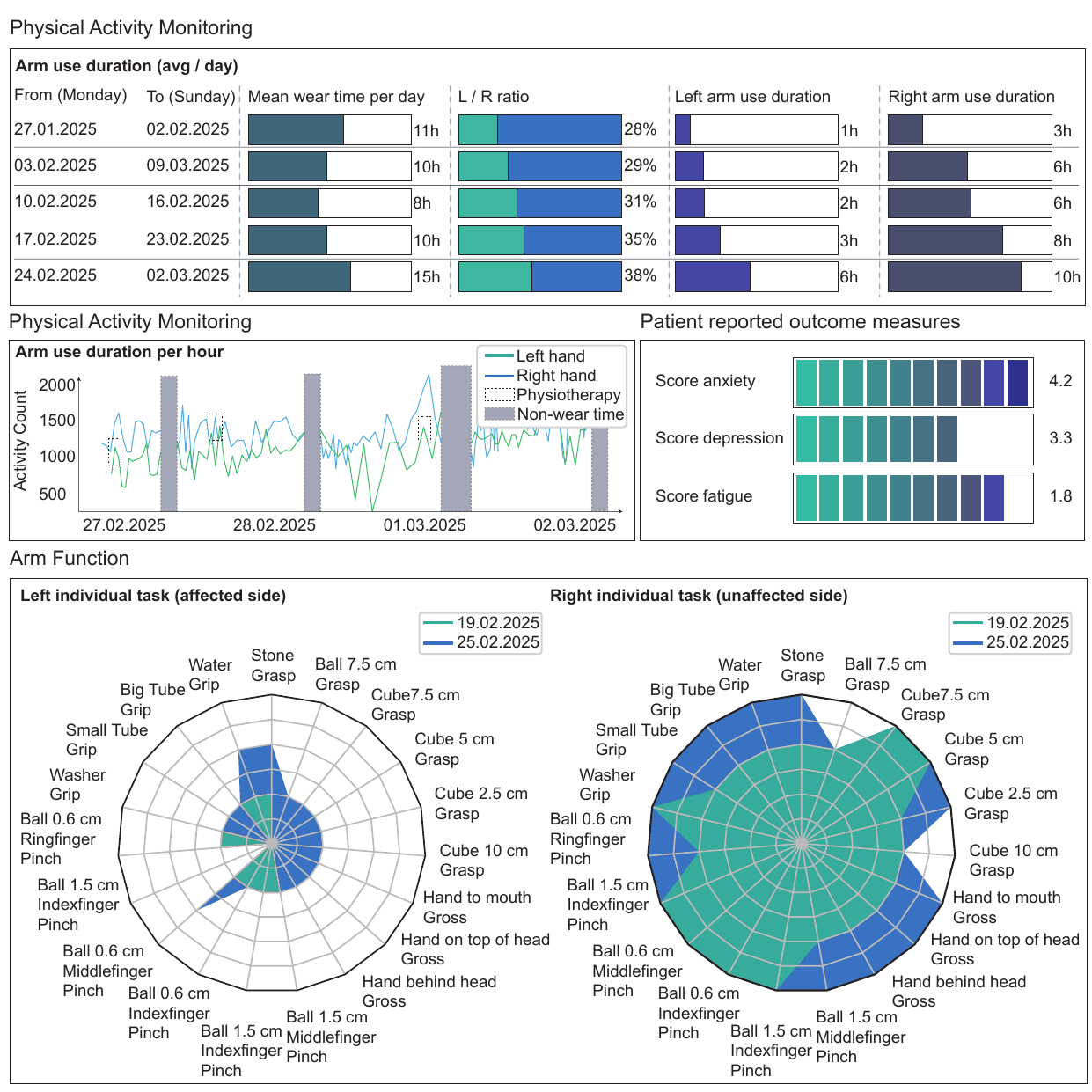}
    \caption{Example visualization of an integrated dashboard embedded via https in a  clinical data collection system. The dashboard is in \textit{clinician perspective} and features representations that were developed iteratively together with the clinical team. Each of the representations allows deeper exploration by selecting it to show more detailed data}
    \label{fig:HC-dash}
\end{figure*}

The data is displayed through a user-friendly Grafana dashboard \cite{grafana} (Figure \ref{fig:HC-dash}), integrated into the clinical information system, where all patient-related data from the clinic is accessible. Each dashboard was designed iteratively using a user-centered approach by a biomedical engineer and clinical expert. Grafana’s flexibility allowed for rapid prototyping and a wide range of visualizations. The clinic’s primary requirement was to visualize clinically meaningful measures in an intuitive and clear manner.

Taken together, our embedded i-health system presents an interactive dashboard to the treating physician, patient, and healthcare staff which visualizes all data acquired from a patient through continuous monitoring, instrumented assessments, Patient Reported Outcomes, intervention logging, and clinical data.

\section{Discussion}

\begin{figure*}[ht]
    \centering
    \includegraphics[width=0.95\textwidth]{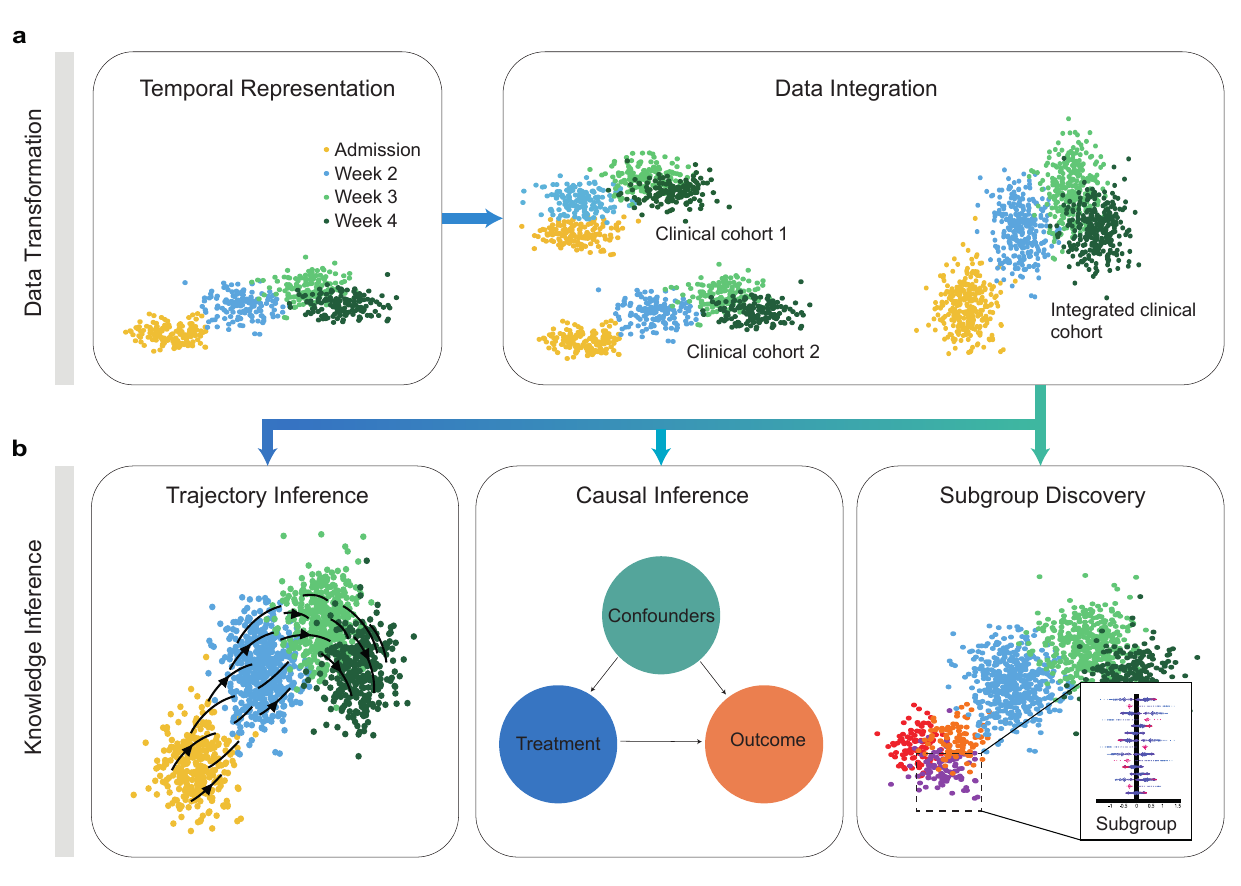}
    \caption{Schematic overview of the temporal representation learning workflow and downstream tasks for time-informed population-level analyses. (a) Multi-modal rehabilitation data, originating from digital assessments, medical records, and interventional data collected across multiple clinical cohorts. (b) Downstream analysis, which depends on the research question and may include causal effect inference, trajectory modeling, or subpopulation analysis.}
    \label{fig:temporal representation}
\end{figure*}

The objective of this work is to create an embedded i-health system for compNR. For this, we designed a blueprint with the requirements of compNR at the center. These include a collection of detailed health status data across clinical and home contexts, native integration of wearable sensors and time-series data, automation of workflows, and inclusion of a secure computational core. As a demonstrator project for the end-to-end process, we customized the i-health system around a stroke rehabilitation clinic, including user centered design principles from the ground up. User needs evaluation indicated a clear requirement for low data capture time, reuse of existing and accepted clinical assessments covering all aspects of the ICF, and user-friendly interfaces.

The implementation of the demonstrator i-health system is largely successful, and it is currently in sustained productive use. Overall, over 1000 patient days have already been captured using the enhanced data collection solution. User satisfaction is high for individual data capture applications, and also high for overall system appreciation. In open interviews across the clinical staff, unfamiliarity with technology, fear of increased workload/complexity, and questionable clinical meaningfulness of novel metrics from the extended data collection were frequently mentioned topics. The main benefits were described as reduced documentation time, accessible data visualization. Furthermore, staff recurrently voiced curiosity concerning the computational aspects of the i-health system. These practical learnings also reflect some of the core questions reported by Lin et al. \cite{lin2024transforming} in their appraisal of requirements for successful implementation of compNR.

Through the implementation, the development team has encountered multiple learnings. Organizational and popular support are paramount for successful integration. Inclusion of the development team into clinical routine processes as observers was important to identify, understand, and contextualize pain-points of the clinical teams. Rapid support, appropriate introductory sessions, and high application reliability were equally central to achieving traction. Overall, our approach fosters a mutually beneficial collaboration between clinics and research institutions, ensuring that data collection is directly linked to patient-centered applications. By co-developing these applications, we not only reduce iteration cycles for new assessments but also enhance clinical practice, improve patient outcomes, and accelerate research translation into clinical routine. Throughout this implementation, we observed a perspective shift as healthcare professionals, previously disengaged from exhaustive data collection for research purposes, began recognizing their value in daily practice. This growing interest is building momentum, transitioning from a research initiative to an integrated hospital service, reinforcing the need to expand and professionalize its offerings to meet clinical demands.

To achieve the necessary impact on delivery efficiency, cost-effectiveness, continuous care, and improved outcomes, future work is expected to focus on developing an analytics framework to continuously model the patient trajectory to allocate optimal treatment interventions. Modeling typically includes an initial patient state, which is captured by the multi-modal data described in this paper, together with a dynamic treatment response. The plastic windows in which treatments are especially effective are short. Hence, healthcare professionals can benefit from some guidance to reduce empirical exploration. One way to approach this is by identifying those former cases most similar to the current patient, together with their therapeutic dynamics. The proposed approach is coupled with the appropriate representation space and explainable machine learning, such that the recommendation of the intervention can be explained to the healthcare professionals. We intend to learn temporal representations from admission onset throughout the rehabilitation journey, including data from different clinical cohorts (Figure \ref{fig:temporal representation}a). We are interested in performing trajectory inference, causal inference and explore the cohorts for clinically meaningful subpopulations, and to determine the features differentially associated with subpopulations \ref{fig:temporal representation}b).
All together, this approach identifies subpopulations by grouping patients based on similarities, examines their NR trajectories, and explores causal relationships between interventions.

\section{Conclusion}
By leveraging a pervasive data-driven approach, this work introduces a scalable and clinically relevant blueprint of an embedded i-health system for compNR. Our transparent and modular three-layer model enables high-fidelity data collection, advanced computational analysis and seamless clinical integration. The collaborative nature of this approach fosters a win-win scenario for research and healthcare, streamlining rehabilitation workflows, and ultimately improving patient outcomes. In a successful real-world implementation, branded as HealthCore, we demonstrate the feasibility of this paradigm shift, paving the way for a future where computational methods play a central role in optimizing NR practice to enhance efficiency and rehabilitation outcomes.

\section*{Acknowledgment}

We would like to thank the clinicians for consulting and support in designing clinical workflows, and research students involved in development and evaluation: Andreas Luft, Meret Branscheidt, Leopold Zizelsberger, Traian Popa, Franziska Riegel, Pranjal Mishra, Dionys Rutz, Alma Flueck, Elena Ruiz, Migjen Shala, Lisa Herndlhofer, Tim Unger, Johannes Pohl, Mathias Bannwart, Anna Schmitt, Leah Schulz, and Loic Kreienbuehl. We thank the patients who generously allowed their data to be reused for this project, and our funders: Open Research Data by the Swiss Data Science Center and the SwissNeuroRehab InnoSuisse Flagship Project (SP3).

\section*{Competing interests}

F.J.T. consults for Immunai, Singularity Bio, CytoReason, Cellarity and Omniscope, and has ownership interest in Dermagnostix and Cellarity. L.H. is an employee of Lamin Labs.

\bibliographystyle{ieeetr}
\bibliography{references, references_benchmark}

\end{document}